\documentclass[showpacs,preprint,amsmath,PRB]{revtex4}
\usepackage{graphicx}
\usepackage{dcolumn}
\usepackage{bm}

\begin{document}

\title{An experimental study on $\Gamma$(2) modular symmetry in the quantum Hall system with a small spin-splitting}
\author{ C. F. Huang $^{1,2}$, Y. H. Chang$^{1*}$, H. H. Cheng$^{3}$, Z. P. Yang$^{3}$, H. D. Yeh$^{1,2}$, C. H. Hsu$^{1}$, C.-T. Liang$^{1}$, D. R. Hang$^{4,5}$, and H. H. Lin $^{6}$ }

\affiliation{$^{1}$Department of Physics, National Taiwan University, Taipei 106, Taiwan, R. O. C.} 
\affiliation{$^{2}$National Measurement Laboratory, Center for Measurement Standards, Industrial Technology Research Institute, Hsinchu 300, Taiwan, R. O. C. }
\affiliation{$^{3}$Center for Condensed Matter Sciences, National Taiwan University, Taipei 106, Taiwan, R.O.C.}
\affiliation{$^{4}$Department of Materials Science and Optoelectronic Engineering, National Sun Yat-sen University, Kaohsiung 804, Taiwan, R.O.C.}
\affiliation{$^{5}$Center for Nanoscience and Nanotechnology, National Sun Yat-sen University, Kaohsiung 804, Taiwan, R.O.C.} 
\affiliation{$^{6}$ Department of Electrical Engineering, National Taiwan University, Taipei 106, Taiwan, R.O.C.}

\date{\today}

\begin{abstract}
Magnetic-field-induced phase transitions were studied with a two-dimensional electron AlGaAs/GaAs system. The temperature-driven flow diagram shows the features of the $\Gamma$(2) modular symmetry, which includes distorted flowlines and shiftted critical point. The deviation of the critical conductivities is attributed to a small but resolved spin splitting, which reduces the symmetry in Landau quantization. [B. P. Dolan, Phys. Rev. B 62, 10278.] Universal scaling is found under the reduction of the modular symmetry. It is also shown that the Hall conductivity could still be governed by the scaling law when the semicircle law and the scaling on the longitudinal conductivity are invalid. \newline
*corresponding author:yhchang@phys.ntu.edu.tw 
\end{abstract}

\maketitle

Magnetic-field-induced phase transitions in two-dimensional electron systems (2DESs) have been an active research topic since the discovery of the quantum Hall effect. [1-18] The law of corresponding states proposed by Kivelson, Lee, and Zhang (KLZ) [2], which was based on the effective field Maxwell-Chern-Simon theory, provides a powerful method for classifying quantum Hall states and the transitions between them. According to the law of corresponding states, all the magnetic-field-induced phase transitions are of an equivalent class. In the integer quantum Hall effect (IQHE), the equivalence is established by the Landau-level addition transformation [1,2]. Magnetic-field-induced phase transitions are believed to be good examples of quantum phase transitions. [1,19] Universal properties such as the reflection symmetry [3], universality of critical conductivities [1,4], and the universal scaling with same critical exponent [5,6] are expected and in addition, it is also expected that the temperature-driven flow lines [1,7,8] are governed by the semicircle law.

Because of the existence of the law of corresponding states, the phase diagram of the QHE has a symmetry equivalent to the $\Gamma _{0}$ (2) symmetry group, which is a subgroup of the modular group. [8-10,20-24] The universal properties mentioned above can be taken as the manifestations of the $\Gamma _{0}$ (2) modular symmetry. [8-10] However, this symmetry relies on the assumption that all the Landau bands are equally spaced in energy, a condition satisfied in the original paper by KLZ, in which the electrons are assumed to be spinless particles. In the presence of a magnetic field, the Zeeman splitting between the spin up and spin down electrons with the same Landau level index is given by $\Delta E = g ^{*} \mu _{B} B$, where $g ^{*}$ is the effective g factor, $\mu _{B}$ is the Bohr magenton, and $B$ is the magnetic filed. [25-27] This energy is usually different from the cyclotron energy of an electron with an effective mass $m ^{*}$, in a magnetic filed $B$, $E=heB/m^{*}$. Therefore, including spin in the problem, the Landau bands become pairs of Landau bands and are no longer all equally spaced. Dolan [9] pointed out that the law of corresponding states proposed by KLZ is suitable when all the Landau bands are well-separated and have no coupling between them. When the spin splitting is small but is resolved, the two Landau bands with the same cyclotron energy are separated only by a small spin gap. The coupling between the two spin states reduce the modular symmetry from $\Gamma _{0}$ (2) to $\Gamma$ (2) [9,10]. With the reduction in symmetry, the particle-hole and Landau-level addition transformations should be modified to [9]
\begin{eqnarray}
\sigma _{xx} (2 - \nu) \leftrightarrow \sigma _{xx} (\nu) \text{, } \sigma _{xy} (2 - \nu) \leftrightarrow 2 e ^{2}/h  -\sigma _{xy} (\nu)  
\end{eqnarray}        
\begin{eqnarray}
\sigma _{xx} ( \nu + 2 ) \leftrightarrow \sigma _{xx} (\nu) \text{, } \sigma _{xy} (\nu +2) \leftrightarrow \sigma _{xy} (\nu) + 2 e ^{2}/h.
\end{eqnarray}
Here $\nu$ is the filling factor, and $\sigma _{xx}$ and $\sigma _{xy}$ are the longitudinal and Hall conductivities, respectively. The main difference between the $\Gamma _{0}$ (2) and the $\Gamma$ (2) symmetries is that although derivation of the semicircle law from the duality is still valid, the critical point in the cross over between two Hall plateaus or between quantum Hall and insulating states is no longer fixed by the duality. In the $\Gamma _{0}$ (2) group, the critical point is fixed at the center of the semicircle in the complex $\sigma$ plane, but the critical point in the $\Gamma$ (2) group can be in any point on the semicircle. In addition, the reduction of the symmetry also results in the distortion of the temperature driven flow diagram. 

In this paper, we report a magnetotransport study on the gated 2DES in an AlGaAs/GaAs heterostructure. We benefit from the fact that this 2DES has been studied before and is well characterized [18]. In this study, a temperature-driven flow diagram with the features due to $\Gamma$(2) symmetry is constructed by investigating a P-P transition between two states separated by small spin-splitting. Universal scaling is observed under the reduced symmetry, and it is found that the scaling on $\sigma _{xy}$ is more robust than that on $\sigma _{xx}$. On the other hand, the feature of $\Gamma _{0}$(2) symmetry is observed in the low-field P-P transition where the spin splitting is unresolved and the transition is between Landau levels with different indices. The sample used in this study was grown by molecular beam epitaxy. The 10 $\times$ GaAs/AlAs ( 1$\times$1 nm) layers, an undoped GaAs layer (500 nm), an undoped Al$_{0.22}$Ga$_{0.78}$As layer as spacer (20 nm), a 40 nm Si-doped Al$_{0.22}$Ga$_{0.78}$As layer with doping concentration 8$\times$10$^{17}$ cm$^{-3}$, and a 10 nm Si-doped GaAs cap layer with the doping concentration $4\times10 ^{17}$ cm$^{-3}$ were grown on top of a semi-insulating GaAs substrate. The sample was made into Hall pattern of 1 mm width with voltage probe spaced 2 mm apart by standard photo-lithography. Indium was alloyed into the contact region at 450 $^{o}C$ in N$_{2}$ atmosphere to form Ohmic contacts, and Al was doposited onto the surface of the Hall bar to form Schottky gate. Magneto-transport measurements were performed with a top-loading He$^{3}$ system and a 0-15 T superconductor magnet. Low frequency AC lock-in technique was used and current of 0.1 $\mu$A was applied to the sample. From Shubnikov-de Haas oscillations and Hall measurement, the carrier concentration $n$ is determined to be $2 \times 10 ^{11} / cm ^{2}$ at $V _{g}=-0.01 V$. At low magnetic fields, the longitudinal resistivity $\rho _{xx}$ increases weakly as the temperature $T$ decreases, and electron mobility $\mu = 7.6 \times 10 ^{3} cm ^{2} / Vs$ is obtained from $\rho _{xx} = 1/ne \mu$ at zero magnetic field at $T$=0.31 K. The mobility is so low that no fractional quantum Hall effect is observed in our study. Therefore, we can focus on the transitions in the IQHE.

Figure 1 shows the curves of the longitudinal and Hall resistivities $\rho _{xx}$ and $\rho _{xy}$ at temperature $T$=0.94 K with $B$=1.5-12 T and gate voltage $V _{g}$=-0.1 V. We can see in Fig. 1 that with increasing $B$, plateaus of filling factor $\nu$=4, 2, and 1 appear successively. Between two adjacent QH plateaus, the 2DES undergoes P-P transitions. Since there is no QH state of the odd filling factor $\nu$=3 between $\nu$=4 and $\nu$=2 QH states, spin-splitting is unresolved in the transition separating $\nu$=4 and 2 QH states. Such a transition is a spin-degenerate P-P transition. On the other hand, the $\nu$=1 QH state due to the spin gap appears at higher $B$, and the P-P transition between $\nu$=2 and 1 QH states is a transition between two closely spaced spin state. As reported in [18], the sample undergoes an insulator-quantum Hall (I-QH) transition at $B=14.7$ T to leave the $\nu=1$ QH state and enter the insulating phase. For convenience, we denote such an I-QH transition as the 1-0 transition, where the number $\lq \lq$0" presents the insulating phase. In the following, we denote P-P transition by the filling factors of its adjacent QH plateaus. Thus we observed the 4-2, 2-1, and 1-0 transitions with increasing $B$. 

Figure 2 shows the curves of the conductivities $\sigma _{xx}$ and $\sigma _{xy}$ obtained from $\rho _{xx}$ and $\rho _{xy}$ by 
\begin{eqnarray}
\sigma _{xx(xy)} = \rho _{xx(xy)} / ( \rho _{xx} ^{2} + \rho _{xy} ^{2} )
\end{eqnarray}
in the 2-1 transition. At $B$=5.49 T, both $\sigma _{xx}$ and $\sigma _{xy}$ are $T$-independent and it is the critical magnetic field, $B _{2-1}$, for the 2-1 transition. The experimental flow diagram for such a transition is shown in Fig. 3, in which each solid line corresponds to the T-driven flow line at a magnetic field and the dash dot line is the trace $\sigma _{xx} ( \sigma _{xy} )$ at $T$=0.31 K, the lowest temperature in our study. In the conventional flow diagram for the
$\Gamma _{0}$(2) group, the transition point is expected to be located at the center of the semicircle, and the $T$-driven flow lines flow in different directions for $B>B _{2-1}$ and $B<B _{2-1}$. In addition, the flow lines are symmetrical with respect to the vertical line passing through the center of the semicircle. However, we could see in Fig. 3 that although the semicircle relation is still valid, as could be seen from the nice semicircle that satisfies the relation
\[
\sigma ^{2} _{xx} + ( \sigma _{xy} -1.5 e ^{2}/h ) ^{2} = ( 0.5 e ^{2} /h ) ^{2},
\]
the transition point, denoted by $Q$ in the figure, is not located at the top of the semicircle. Although the flow lines at both sides of $Q$ do flow to opposite directions, starting from point $A$, at which $\sigma _{xy} > 1.5 e ^{2} / h$, the flow line moves toward to the point ($e ^{2} /h$,0) rather than ($2e ^{2}/h$,0) with decreasing $T$, namely, the temperature-driven flow lines are distorted. Clearly, the 2-1 transition we observed has the properties of the $\Gamma$(2) group: valid semicircle relation, system-dependent transition point, and distorted flow diagram.

At low $T$, magnetic-field-induced phase transitions are expected to follow the two-parameter scaling such that both $\sigma _{xx}$ and $\sigma _{xy}$ (or $\rho _{xx}$ and $\rho _{xy}$) are functions of the scaling parameter. [5,6,9,10] Figure 4 shows the curves of $\sigma _{xx}$ and $\sigma _{xy}$ at different temperatures with respect to the scaling parameter $s \equiv ( \nu- \nu _{2-1} ) / T ^{\kappa}$ with $\kappa=0.4$, where $\nu _{2-1}$ is the filling factor at $B _{2-1}$. We could see in this figure that the curves of $\sigma _{xy}$ at different $T$ collapse into a single curve and hence $\sigma _{xy}$ is a function of the scaling parameter in the whole transition region. However, we found that although the curve $\sigma _{xx}$ collapse nicely for $\nu > \nu _{2-1}$, they do not merge together for $( \nu - \nu _{2-1} ) / T ^{\kappa} <-0.017$. When the scaling on $\sigma _{xx}$ fails, we found that the $T$-driven flow lines move away from the semicircle with increasing $T$. For example, the flow line reaches point $C$ at $T$=0.94 K in Fig. 3. It is shown that in the effective Lagrangian that determines the properties at the critical points, $\sigma _{xy}$ and $\sigma _{xx}$ are coefficients of the topological and kinetic terms, respectively. [20,28] The topological term is an important quantity to construct the phase diagram of the quantum Hall effect [28,29], and it has been shown that the critical point in $\sigma _{xy}$ is more robust than that in $\sigma _{xx}$ with respect to the temperature $T$ [30]. Our study shows that the universal scaling on $\sigma _{xy}$ can remain correct when the $T$-driven flow lines do not follow the semicircle law and the scaling on $\sigma _{xx}$ is invalid, and supports the claim that $\sigma _{xy}$ is more robust than that in $\sigma _{xx}$ with respect to the temperature $T$ [30]. 

With increasing $B$, as reported in Ref. [18], we observed the 1-0 transition which follows the universal scaling and semicircle law while the resistivitiy at the crtical field $B$=14.7 T is not universal. The critical magnetic field is close to the maximum field 15 T of our system, so we cannot investigate the details about the insulating phase such as whether it is terminated by a fractional quantum Hall liquid at higher $B$ or not. [31,32] But the scaling behaviors near $B$=14.7 T show that we do observe the 1-0 transition. The $T$-driven flow lines near its critical magnetic field are shown in Fig. 5, in which the flow lines are of opposite directions near the point $Q ^{\prime}$. Such a point corresponds to the critical point in the 1-0 transition and serves the same role as the point $Q$ in the 2-1 transition. As shown in the inset of Fig. 5, the curves $\sigma _{xx} ( \sigma _{xy} )$ at $T$=0.31-0.94 K are along the expected semicircle because of the semicircle law. The point $Q ^{\prime}$ deviates from the top position of the semicircle, and the flow diagram also shows the features of $\Gamma$(2) symmetry. The deviation on such a point, however, might either be due to the inappropriate conversion form resistivities to conductivites or the reduction of modular symmetry. [9] In the 1-0 transition reported in Ref. [33], the features of $\Gamma$(2) symmetry are removed by renormalizing $\rho _{xx}$ [9,34]. To further investigate the 1-0 transition, it is noted that the particle-hole transformation is generalized as in Eq. (1) if the modular symmetry is reduced to $\Gamma$ (2). Under the generalized particle-hole transformation, the flow diagrams for the 2-1 and 1-0 transitions are symmetric with respect to the vertical line $\sigma _{xy} = e ^{2} /h $. [9] We can see in Figs. 5 and 3 that $\sigma _{xy} < 0.5 e ^{2} /h$ at $Q ^{\prime}$ while $\sigma _{xy} > 1.5 e ^{2} /h$ at $Q$, and the value of $\sigma _{xx}$ are close at these two points. Therefore, the flow diagrams have the feature of generalized particle-hole transformation, which indicates that the 1-0 transition follows $\Gamma$(2) symmetry.

Using the dissipation equation [35-38], we estimated that the spin splitting in our sample is about 0.04 of the cyclotron energy from the value of $\rho _{xx}$ at $T$=0.94 K and 0.68 K at $B$=9.8 T, where $\rho _{xx}$($B$) reaches its minimum in the $\nu$=1 QH state. Such a value is larger than the ratio of the bare spin gap to cyclotron splitting in the 2D GaAs system, which indicates the importance of the exchange enhancement on Zeeman spin splitting. [25-27] In Ref. [1], Shahar et al. also reported a study on the 2-1 and 1-0 transitions in 2DES in an AlGaAs/GaAs heterostructure. In their report, the carrier concentration and mobility of the sample are $n=2.7 \times 10 ^{11} /cm ^{2}$ and $\mu =1.08 \times 10 ^{4} cm ^{2}/Vs $. Although both $n$ and $\mu$ of their sample are similar to those in our sample, the modular symmetry is not reduced from $\Gamma _{0}$(2) to $\Gamma$(2) in their results. It should be noted that the effective spin splitting for QH systems may be sample dependent. [39] Hence the different modular symmetries might show up for sample with different effective spin gaps. In addition, it is in debate whether the dissipation equation can yield the spin gap. [40-42] More studies are necessary to see how to obtain the effective gap to determine the modular symmetry quantitatively.

To study the reduction of modular symmetry due to the small spin gap, the spin splitting has to be resolved enough to separates Landau bands. On the other hand, the splitting also has to be so small that there exists coupling between two Landau bands separating by a spin gap. If the splitting is so large that the coupling can be ignored, we should observe spin-resolved transitions governed by $\Gamma _{0}$(2) symmetry just as in the report of Shahar et al. [1]. But if the spin splitting is too small and is unresolved, all (spin-degenerate) Landau bands are evenly separated and there should be no reduction of modular symmetry. Therefore, it is uneasy to observe $\Gamma$ (2) symmetry. At low fields, as mentioned above, there is no QH state of the odd filling factor in our study and thus the spin splitting is unresolved. In Fig. 1 we could observe the 4-2 transition, which is a spin-degenerate transition. At the critical field $B _{4-2}$ of such a transition, $\rho _{xx}$ and $\rho _{xy}$ should equal $0.1 h/e ^{2}$ and $0.3 h/e ^{2}$ if the semicircle law is vald and the unstable point in the $\sigma _{xy}$-$\sigma _{xx}$ plane is at the top of the semicircle. Although in the $\nu$=4 QH state $\rho _{xx}$ does not approach zero in our study, the quantum interference leading to QH states can still induce plateaus in $\rho _{xy}$. [43] In our study, we can identify $B _{4-2}$=2.46 T at which $\rho _{xy}$ is $T$-independent while in $\rho _{xx}$ there is no well-defined $T$-independent point. In Fig. 6, at $B _{4-2}$ the Hall resistivity $\rho _{xy}$ = 7.75 $\pm$ 0.05 k$\Omega$ is close to the expected value of 0.3 $h/e ^{2}$. Therefore, the universality of critical Hall resistivity is valid in such a spin-degenerate transition, and the symmetry is the $\Gamma _{0}$(2) symmetry. 

To further investigate the modular symmetry in the gated 2DES, we change the gate voltage $V _{g}$ from -0.1 V and 0V. Well-developed spin-resolved P-P transition is observed between QH states of $\nu$=2 and 1, and the reduction of the modular symmetry can still be identified in such a transition by investigating the $T$-driven flow lines and critical points. In addition, the universal scaling is also observed in $\sigma _{xy}$ even when the semicircle law and the scaling on $\sigma _{xx}$ both become invalid. On the other hand, the critical value of $\rho _{xy}$ in the 4-2 transition maintains at 0.3 $h/e ^{2}$ when $V _{g}$ is changed, which indicates there is no reduction of modular symmetry in such a spin-degenerate transition. Thus we observe the reduction of modular symmetry from $\Gamma _{0}$ (2) to $\Gamma$ (2) again at $V _{g}$=0 V. 

In conclusion, we performed a magnetotransport study on a gated 2DES in an AlGaAs/GaAs heterostructure to study the universality of magnetic-field-induced phase transitions. A temperature-driven flow diagram is constructed by studying the 2-1 transition, and our study supports Dolan's suggestion [9] that we shall consider $\Gamma$(2) symmetry under a small but resolved spin splitting, which breaks $\Gamma _{0}$(2) symmetry. The reduction of the symmetry from $\Gamma _{0}$(2) to $\Gamma$(2) is due to the unequal energy splittings arising from Zeeman effect and Landau quantization. We also showed that the energy splitting is important to the scaling on $\sigma _{xx}$ and semicircle law over the same temperature range. The modular symmetry could also be reduced in the 1-0 transition since its critical point can be related to that of the 2-1 transition by the generalized particle-hole transformation. The universal scaling on $\sigma _{xy}$, in fact, can survive when the scaling on $\sigma _{xx}$ is invalid and the temperature-driven flow lines are not along the expected semicircle. 

This work is supported by the National Science Council and Ministry of Education of the Republic of China. D. R. Hang acknowledges support from Aim for the Top University Plan, ACORC, NSYSU. One of the authors (C. F. H.) thanks B. P. Dolan and C. P. Burgess for valuable discussions.

\newpage

Fig. 1  The longitudinal and Hall resistivities $\rho _{xx}$ and $\rho _{xy}$ at the temperature $T$=0.94 K when the magnetic field $B$=1.5-12 T and the gate voltage $V _{g}$=-0.1 V. \newline  
Fig. 2  Traces of the longitudinal and Hall conductivities $\sigma _{xx}$ and $\sigma _{xy}$ in the 2-1 transition. The dash-dot line labels the critical point at the field $B _{2-1}$=5.49 T. \newline
Fig. 3  The temperature-driven flow diagram for the 2-1 transition. The dash dot line is the curve $\sigma _{xx} ( \sigma _{xy} ) $ at $T$=0.31 K and each solid line corresponds to a temperature-driven flow line at a magnetic field. The symbols boxes, diamonds, triangles, circles are for the points at $T$=0.94 K, 0.68 K, 0.49 K, and 0.31 K, respectively. The point $Q$ corresponds to the critical point of the 2-1 transition. We could see that the flow line starting at the point $A$ does not move toward the point (2$e ^{2}/h$,0) as $T$ decreases. At the left hand side of the point $Q$, there are flow lines away from the semicircle at higher temperatures, e.g., the flow line starting from the point $C$. The dash-dot line and the solid line in the inset are the curve $\sigma _{xx} (\sigma _{xy})$ at $T$=0.31 K and the semicircle defined by Eq. (2), respectively. \newline
Fig. 4  Curves of $\sigma _{xx}$ and $\sigma _{xy}$ with respect to the scaling parameter $(\nu-\nu _{2-1} )/T ^{\kappa}$ ,where the critical exponent $\kappa=0.4$, $\nu$ is the filling factor, and $\nu _{2-1}$ is the filling factor at the critical point. \newline
Fig. 5  The temperature-driven flow diagram for the 1-0 transition near the critical point. The dash dot dot line is the semicircle satisfying $\sigma _{xx} ^{2} + ( \sigma _{xy} - 0.5 e ^{2} / h ) ^{2} = (0.5e ^{2}/h) ^{2}$ and each solid line corresponds to a temperature-driven flow line at a magnetic field. The symbols boxes, diamonds, triangles, circles are for the points at $T$=0.94 K, 0.68 K, 0.49 K, and 0.31 K, respectively. The point $Q ^{\prime}$ corresponds to the critical point of the 1-0 transition. In the inset $\sigma _{xx} (\sigma _{xy})$ at $T$=0.94 K, 0.68 K, 0.49 K, and 0.31 K, respectively, and the dash dot dot line is the semicircle. \newline
Fig. 6  The curves of $\rho _{xy}$ in the 4-2 transition. The dash dot line labels the critical point magnetic field $B _{4-2}$=2.46 T.

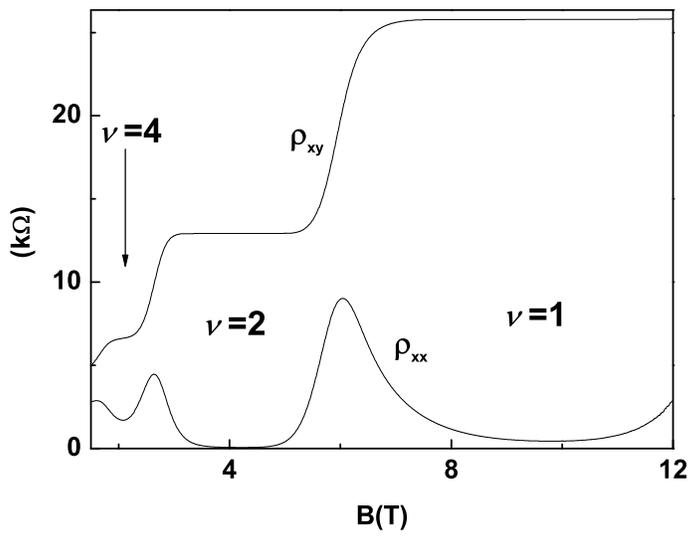
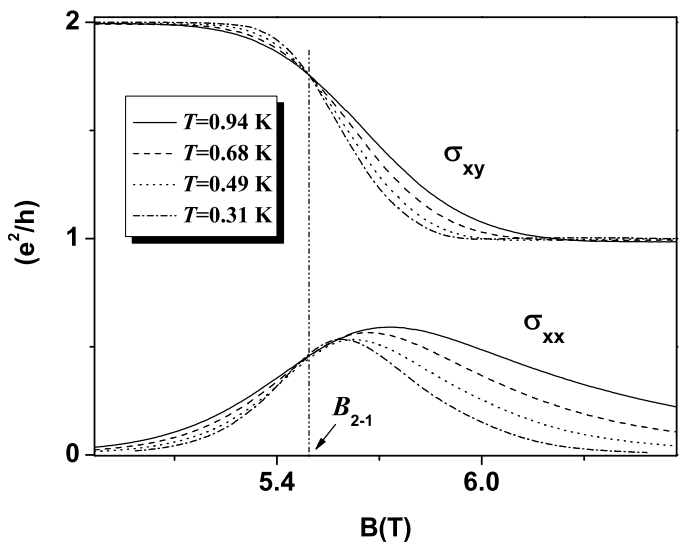
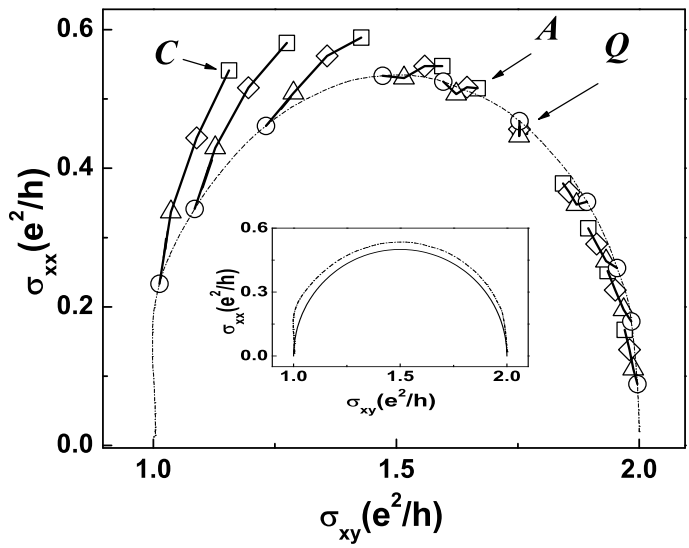
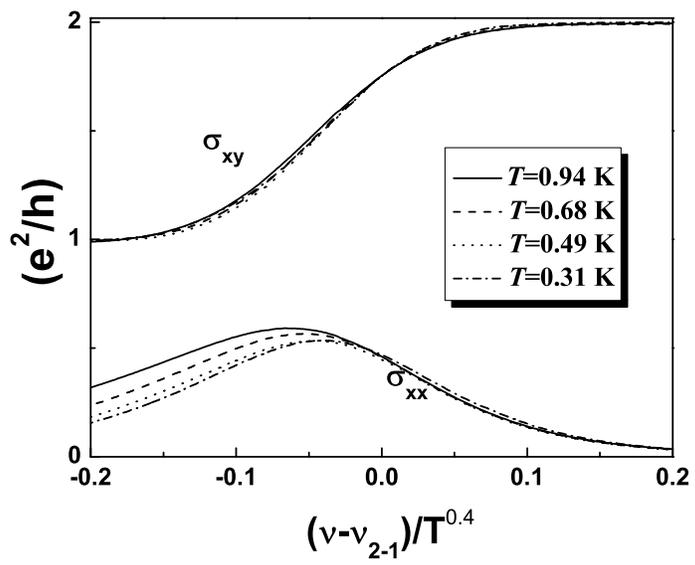
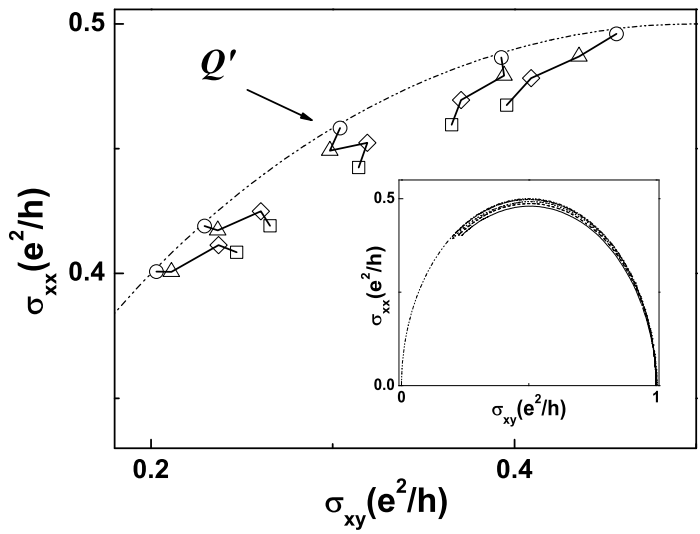
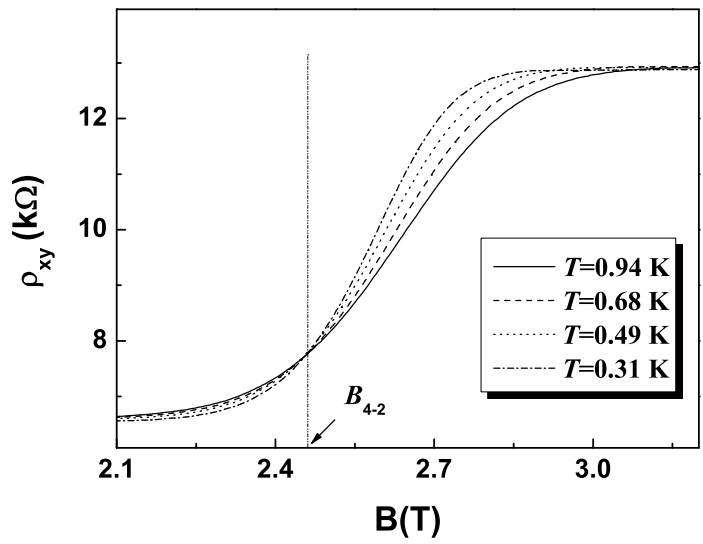
\end{document}